\documentclass[conference,compsoc]{IEEEtran} 
\usepackage{graphicx} 
\usepackage[ruled, lined, linesnumbered, commentsnumbered, longend]{algorithm2e}
\usepackage{multirow}
\usepackage{pifont}
\usepackage{subcaption} 
\usepackage{booktabs}
\usepackage{amsmath,amsthm,amssymb}

\usepackage[T1]{fontenc}
\usepackage[hidelinks]{hyperref}
\usepackage{cleveref}
\usepackage{url}

\usepackage{array}
\newcolumntype{L}[1]{>{\raggedright\let\newline\\\arraybackslash\hspace{0pt}}m{#1}}
\newcolumntype{C}[1]{>{\centering\let\newline\\\arraybackslash\hspace{0pt}}m{#1}}
\newcolumntype{R}[1]{>{\raggedleft\let\newline\\\arraybackslash\hspace{0pt}}m{#1}}

\newcommand{\xx}{{\bf x}}
\newcommand{\xxx}{X}
\newcommand{\yy}{{\bf y}}
\newcommand{\yyy}{Y}

\newcommand{\sZ}{\mathbb{Z}}

\newcommand{\sX}{\mathcal{X}}

\newcommand{\name}{\textsc{ZK-Series}}
\newcommand{\Sharp}{\textsc{Sharp}}

\newcommand{\cmark}{\textcolor{green!60!black}{\ding{51}}}
\newcommand{\xmark}{\textcolor{red!60!black}{\ding{55}}}

\usepackage{xargs}
\usepackage[pdftex,dvipsnames]{xcolor} 
\usepackage[colorinlistoftodos,prependcaption,textsize=tiny]{todonotes}
\newcommandx{\commentt}[2][1=]{\todo[linecolor=red,backgroundcolor=red!25,bordercolor=red,#1]{#2}}

\begin{document}

\title{\name{}: Privacy-Preserving Authentication using Temporal Biometric Data}

\author{
    \IEEEauthorblockN{Dani\"el Reijsbergen, Eyasu Getahun Chekole, Howard Halim, Jianying Zhou} \\
    \IEEEauthorblockA{Singapore University of Technology and Design
    \\\{daniel\_reijsbergen,eyasu\_chekole,howard\_halim,jianying\_zhou\}@sutd.edu.sg}
}

\maketitle

\begin{abstract}
    \textit{Biometric} authentication 
    relies on physiological or behavioral
    traits that are inherent to a user,
    making them difficult to lose, forge or forget.
    Biometric data with a \textit{temporal} component enable the following authentication protocol:
    recent readings of the underlying biometrics are encoded as \textit{time series} and compared to a set of \textit{base} readings. If the \textit{distance} between the new readings and the base readings
     falls within an acceptable threshold, then the user is successfully authenticated. Various methods exist 
     for comparing time series data,
     such as Dynamic Time Warping (DTW) and the Time Warp Edit Distance (TWED), 
     each offering advantages and drawbacks 
     depending on the context. Moreover, many of these techniques
     do not inherently preserve privacy, which is a critical consideration in biometric authentication due to the complexity of resetting biometric credentials. 
    
    In this work, we propose \name{} to provide privacy and efficiency to a 
    broad spectrum of time series-based authentication protocols. \name{} uses the same 
    building blocks, i.e., zero-knowledge multiplication proofs and efficiently batched range proofs, 
    to ensure consistency across all protocols. Furthermore, it is optimized for compatibility with low-capacity devices such as smartphones. To assess the effectiveness of our proposed technique, we primarily focus on two case studies for biometric authentication: shake-based and blow-based authentication. To demonstrate \name{}'s practical applicability even in older and less powerful smartphones, we conduct experiments on a 5-year-old low-spec smartphone using real data for two
     case studies 
    alongside scalability
    assessments using artificial data. Our
    experimental results indicate that the privacy-preserving authentication protocol can be
    completed within 1.3 seconds on older devices.
\end{abstract}

\section{Introduction}
\label{sec:introduction}

Authentication has become an everyday activity for most individuals in a modern society, e.g., to unlock a smartphone or access a banking app. 
While the frequency with which users log into their devices has increased, so have the stakes: for example, an attacker gaining wrongful access to a banking app may result in large financial losses. As such, there is a simultaneous need to improve the convenience \textit{and} security of authentication, and these two goals can be contradictory \cite{karapanos2015sound}.
Recently, biometric protocols (e.g., voice or face recognition) have emerged as an alternative to knowledge-based (e.g., passwords) and possession-based (e.g., a hardware security token) protocols \cite{meng2014surveying,alzubaidi2016authentication}. Biometric authentication has good performance in terms of convenience and security because biometric data is inherent to the user and cannot be lost or forgotten \cite{meng2014surveying}.
However, the leakage of the underlying biometric data is a considerable threat because such data is
i) prohibitively difficult to revoke and reset \cite{patel2015cancelable}, and
ii) increasingly easy to replicate due to the rise of generative AI and deepfakes \cite{salko2024security}.
To mitigate this threat, there is increasing interest in \textit{soft} or \textit{behavioral} biometrics \cite{lien2023challenges} (e.g., eye movements, breath patterns, and gait) that rely on users' inherent traits while being easier to reset than \textit{hard} biometrics (e.g, facial recognition, iris scans, and fingerprint verification).
A downside of soft biometrics is that they typically have lower authentication accuracy than hard biometrics \cite{bailey2014user}, and although they are easier to revoke than hard biometrics, privacy remains paramount.

Our main goal is to design a practical protocol for user authentication using soft biometrics. Our primary use case is for users to log into a critical smartphone app, e.g., for online banking. In this case, the authentication protocol is implemented through a separate app\footnote{Existing authentication apps such as Google Authenticator or SingPass do not support (soft) biometrics.} that sends authentication proofs to the critical secondary app. Any practical authentication protocol must, at least, satisfy the following five requirements. The first requirement is \textit{generality} in the sense that it supports techniques for soft biometric data with a temporal (time series) component, e.g., gait or stroke dynamics. For such data, techniques such as Dynamic Time Warping (DTW) are well-known to have high accuracy \cite{belkhouja2022dynamic,keogh2005exact}. The second is \textit{security} against attackers who have full access to the user's device and who may produce deepfake signals, but who do not have access to the user's biometric data. The third is \textit{privacy}, i.e., the protocol does not reveal raw biometric data to the (possibly malicious) secondary app. The fourth is that all computations are performed \textit{locally}, so that no raw data ever leaves the user's device. The fifth and final requirement is \textit{efficiency}, which means that all computations can also be performed on low-end devices such as older smartphones within seconds. 

Existing approaches violate at least of one of these conditions. For example, the BioZero protocol \cite{lai2024biozero} does not satisfy generality as it does not support DTW, whereas other works \cite{liu2019privacy,zhu2014privacy,zheng2021efficient} that support privacy-preserving protocols for DTW comparison have computation times (over 200s, 18s, and 6.8s for a single pair of time series, respectively) that do not satisfy our efficiency requirement. Research efforts that add privacy and revocability to hard biometrics such as fingerprint scans, e.g., through the use of fuzzy vaults \cite{juels2006fuzzy,sarkar2020review}, are orthogonal to this work.
Soft biometrics can be used in conjunction with hard biometrics for additional security, or when conditions render such hard biometric techniques impractical (e.g., when light conditions do not allow for face or iris scans).

In this work, we propose the \name{} authentication protocol to achieve all of the above requirements.
To achieve generality, \name{} introduces a broad class of authentication protocols (\Cref{sec:proof_structure}) consisting of three components: local distance (e.g., Euclidean), series distance (e.g., DTW), and authentication method (e.g., distance threshold to $k$ nearest neighbors). This class is broader than achieved by other works, as illustrated in  \Cref{tab:protocols_overview}. 
To achieve security and privacy while maintaining practical performance, \name{} leverages a recent development in zero-knowledge range proofs, i.e., \Sharp{} range proofs \cite{couteau2022sharp}, that enables efficient proof batching. This approach has  lower average proof generation times \cite{christ2024sok} than more established  approaches such as Bulletproofs \cite{bunz2018bulletproofs} or Groth16 \cite{groth2016size}, which are designed for a setting with high-capacity provers and low-capacity verifiers. Low proof generation times are a key advantage in our setting, as both proof generation and verification typically occur on a low-capacity device such as a smartphone. 

We illustrate the advantages of \name{} through two case studies in which i) shake patterns and ii) blow acoustics are used for authentication. Both datasets have been made available via \cite{zkseries_data}. The first dataset consists of time series data from 20 participants 
where each user generated 10 base readings of four different phone sensors (gyroscope,
orientation, accelerometer, and magnetometer) and three coordinates per sensor. The second dataset 
includes data from 50 participants and consists of time series recordings of 
10 distinct phone-blowing acoustic patterns, each performed over a period of 5 seconds.
Our experiments suggest that the Manhattan distance and DTW achieve noticeably higher accuracy than the Euclidean sum, thus illustrating the need for a framework with greater protocol support than related work such as BioZero \cite{lai2024biozero}. Even if the 3-dimensional input signal of a single sensor is used for the shake-based case study, \name{} can achieve a false positive rate of less than 0.1\%, and this signal can be combined with other biometrics such as face or voice scans for greater accuracy. 
The entire protocol takes around 1.3 seconds to complete on a Samsung Galaxy A71 smartphone from 2019, which has an 8-core CPU. This indicates that \name{} is suited for practical use in low-spec smartphones, and an improvement on the 2 seconds needed by BioZero on a 10-core CPU. 
We code for our experiments has been made available via \cite{zkseries_code}.

\setlength{\tabcolsep}{2pt}
\begin{table}[b]
\caption{Comparison between \name{} and other works in terms of protocol support, privacy, and practicality.}
\label{tab:protocols_overview}
\begin{tabular}{llccccc}
\multicolumn{2}{l}{\hspace{0.4cm}protocol support} & \cite{liu2019privacy}& \cite{zhu2014privacy} & \cite{zheng2021efficient} & \cite{lai2024biozero} & \textbf{\name{}} \\ \toprule \toprule

\multirow{3}{*}{local} & Manhattan & \cmark & \cmark & \cmark & \xmark & \cmark \\
& (squared) Euclidean  & \cmark & \cmark & \xmark & \cmark & \cmark \\
& Chebyshev & \xmark & \xmark & \xmark & \xmark & \cmark \\ \midrule

\multirow{4}{*}{series\;\;} & sum & \xmark & \xmark & \xmark & \cmark & \cmark \\
& DTW & \cmark & \cmark & \xmark & \xmark & \cmark  \\
& discrete Fr\'echet   & \xmark & \cmark & \xmark & \xmark & \cmark  \\
& TWED  & \xmark& \xmark & \cmark & \xmark & \cmark  \\ \midrule

\multirow{3}{*}{auth} & nearest + threshold & \xmark & \xmark & \xmark & \cmark & \cmark \\
 & $k$ nearest + sum thres. & \xmark & \xmark & \xmark & \xmark & \cmark \\
 & $k$ nearest + max.\ thres. & \xmark & \xmark & \xmark & \xmark & \cmark \\ \midrule \midrule
 \multicolumn{2}{l}{privacy} & \cmark & \cmark & \cmark & \cmark & \cmark \\
 \multicolumn{2}{l}{practical efficiency} & \xmark & \xmark & \xmark & \cmark & \cmark \\\bottomrule \bottomrule

\end{tabular}
\end{table}
\setlength{\tabcolsep}{5pt}

\subsection*{Contributions}

In summary, our contributions are as follows.
\begin{itemize}
    \item We present \name{}, which enables privacy-preserving authentication for biometric data with a temporal (time series) component.
    \item \name{} supports a wide range of advanced time series comparison protocols: its generality exceeds related works \cite{lai2024biozero,liu2019privacy,zhu2014privacy,zheng2021efficient} on three levels: more local distance metrics, more time series distance functions, and authentication protocols beyond a single 1-to-1 comparison.
    \item We present a detailed performance evaluation based on two case studies involving shake-based and blow-based authentication. We find that, unlike \cite{liu2019privacy,zhu2014privacy,zheng2021efficient}, \name{} has practical performance  on a five-year-old low-spec smartphone.
\end{itemize}
\emph{Organization.} This paper is structured as follows. We first discuss related works in \Cref{sec:preliminaries}.
We then discuss our system and threat model in \Cref{sec:system_model} and the \name{} protocol in Sections~\ref{sec:our_proposal}. We discuss our proof-of-concept implementation in \Cref{sec:implementation}, and present our performance evaluation in \Cref{sec:evaluation}. We conclude the paper in \Cref{sec:conclusions}.

\section{Related Work}
\label{sec:preliminaries}
BioZero \cite{lai2024biozero} is the authentication protocol that is most closely related to \name{}. BioZero authenticates users by comparing a single base reading of their biometric data to a recent reading.
In particular, the sum of squared (Euclidean) differences between each point in the two readings is calculated, and if the difference is small enough, then a zero-knowledge proof of this statement is computed through the Groth16 zk-SNARK \cite{groth2016size}. 
Various other works have been proposed to provide privacy to time series analysis. Liu et al.\ present a DTW-based approach that uses Yao's garbled circuits, a form of secure multiparty computation (MPC), to ensure data privacy \cite{liu2019privacy}. They consider a setting in which two independent healthcare providers want to determine the distance between two time series without revealing the underlying values to each other. The main use case is one where a provider with one reading seeks to find the closest reading among a set held by another provider. As with many approaches based on MPC, this approach is hampered by low efficiency: the communication overhead for determining DTW metrics is in the order of gigabytes, and the duration of a single DTW call on two series of length 128 is more than 200s. 
Zhu et al.\ \cite{zhu2014privacy} also provide an early MPC-based method of obtaining the DTW distance between two time series -- again, performance is an obstacle, with the protocol requiring more than 18s to compute the DTW between two time series of length 100. In addition to DTW, they also support the discrete Fréchet distance. Zheng et al.\ \cite{zheng2021efficient} propose a privacy-preserving method to compare a time series to a large set of base readings using TWED: as mentioned before, a key advantage of TWED is its triangle inequality which enables indexing and pruning. In \cite{zheng2021efficient}, the ability to prune the dataset reduces computation times by $97.9\%$. Without pruning, their method takes $95,912$s on 14,000 series with $m=10$ and $T=96$, so roughly $6.8$s per time series on a PC.

Wang et al.\ propose to augment the privacy of time series data storage using differential privacy \cite{wang2020towards} -- however, this is not applicable for authentication because it still reveals the (distorted) base readings, which makes it susceptible to false data injection.
Zheng et al.\ \cite{zheng2021efficient} propose a method based on symmetric homomorphic encryption to perform privacy-preserving range queries on time series data using the TWED. There are some similarities with authentication, i.e., if authentication is successful if the distance to the closest match is small enough. However, performance is again an obstacle, with verification times of roughly 15 seconds for time series of length 60.

Fuzzy vaults, originally proposed by Juels and Sudan \cite{juels2006fuzzy} and subsequently implemented in a wide range of contexts \cite{sarkar2020review}, achieve both revocability and privacy for biometric authentication techniques. In this approach, a secret is encoded as a polynomial, and the base readings are used as inputs for this polynomial. This results in a series of genuine points, which are then obfuscated using random `chaff' points. To authenticate, this procedure is executed on a new reading and authentication succeeds if the overlap with the original points is substantial. Revocation can be achieved by switching to a different secret. Although fuzzy vaults tolerate some variability in the input data, high levels of intra-user variability still cause performance to degrade: e.g., for fingerprint biometrics, which have limited variability, fuzzy vaults typically achieve low false acceptance rates \cite{sarkar2020review}, but performance is worse (above $2\%$ false acceptance rate) for electronically drawn signatures \cite{bhateja2014robust}. Since soft biometrics such as shake-based and blow-based signals typically exhibit high variability, we do not consider this approach further.

\section{System and Threat Model}
\label{sec:system_model}
\subsection{System Entities}
\label{sec:system_entities}

\name{} is designed for a system with the following entities:

\textbf{User.} The user's goal it to authenticate to gain access to a secure resource, e.g., a banking app or a location secured by an external device. 

\textbf{User Device.} The user has access to a handheld device with moderate computational power, e.g., a smartphone.

\textbf{Sensors.} The sensors generate readings of the user's biometric \textit{input signals}, e.g., audio, video, touch, or motion. These sensors are either present on the user's device (e.g., the phone's gyroscope), an external device (e.g., a security camera), or both.

\textbf{\name{} App.} The \name{} app runs on the user's handheld device and stores the user's raw biometric data. Furthermore, the \name{} app performs computations on the raw data and generates zero-knowledge proofs to show that the computations were performed correctly.

\textbf{Secondary App.} The secondary app's goal is to prevent unauthorized access by an attacker to the secure resource. It either runs on the user's device, or on another device -- e.g., a security system encompassing a processor and sensors at a restricted location.

\textbf{Bulletin Board.} Cryptographic commitments to the base readings are stored on a publicly accessible, tamper-evident bulletin board, e.g., a trusted server or a public blockchain. This allows the secondary app to verify that the base readings are kept consistent between authentication attempts. 

\subsection{Data Model}
\label{sec:data_model}

\textbf{Notation.} In the following, we write $\sZ$ for the set of all integers, $\sZ_+$ for the nonnegative integers, and $\sZ_p =\{0,\ldots,p-1\}$ for any $p\geq1$. 
For any finite set or sequence $S$, we denote the number of elements by $|S|$. We do not consider operations on non-integer values (e.g., floating point numbers) because the cryptographic techniques of \Cref{sec:crypto_primitives} are only defined on groups of integers modulo a (large) prime. A summary of the key notation presented in this section can be found in \Cref{tab:notation}.

\begin{table}[b]
    \centering
    \caption{List of symbols.}
    \begin{tabular}{C{0.3\linewidth}|L{0.6\linewidth}}
        symbol & meaning \\ \toprule
         $m$ & dimension of each time series element \\  \midrule
         $\sZ^m$ & set of all m-dimensional vectors whose elements are values in $\sZ = \{0,1,\ldots\}$ \\ \midrule
         $\sX$ & set of all time series whose points are elements of $\sZ^m$ \\ \midrule
         $d : \sZ^m \times \sZ^m \rightarrow \sZ$ & local distance between two elements in a time series \\ \midrule
         $\delta : \sX \times \sX \rightarrow \sZ$ & series distance between two time series \\ \bottomrule
    \end{tabular}
    \label{tab:notation}
\end{table}

\textbf{Time Series.} Authentication protocols enabled by \name{} operate on sets of time series $\{X_n\}_{n=1,\ldots,N}$. Each time series in such a set is denoted by \mbox{$X = (\xx_1,\ldots,\xx_{T})$}, where the index represents time and $T = |X|$ the length of the time series. 
Let $\sX$ denote the set of all time series. Each time series element $\xx_t \in \sZ_p^m = (x_{t,1},\ldots,x_{t,m})$ represents a measurement of $m$ input signals, where $x_{t,j}$ denotes the measurement of input signal $j$ at time $t$. 
The nature of the input signals depends on the sensors and their devices.
A modern smartphone will typically have native support for converting input signals into floating point numbers, which can then be converted into (positive) integers through normalization and rounding. 
Other signals such as audio and video can be converted to multi-dimensional time series through feature extraction, but a more detailed discussion is beyond the scope of this work.

Unlike other works \cite{marteau2008time,chen2004marriage}, we do not consider \textit{gaps} in time series data, which may occur due to, e.g., measurement errors or devices going offline. The reason is that, in our setting, users can be asked to re-record measurements when sensors temporarily fail. Additionally, some works consider timestamp differences between time series elements \cite{marteau2008time} -- in our setting, the app records at preset times, so we assume that the time differences between elements at the same time index in different time series are negligible.

\textbf{Distance Functions.} The distance between two time series can be expressed via two functions: the \textit{local} distance and the \textit{series} distance. The local distance is denoted by a function $d:\sZ^m \rightarrow \sZ_+$. Common examples of local distance functions include:
\begin{itemize}
	\item $d_1(\xx_t,\yy_{t'}) = \sum_{j=1}^m |x_{t,\,j}- y_{t',\,j}|$,
	\item $d_2(\xx_t,\yy_{t'}) = \sum_{j=1}^m (x_{t,\,j}- y_{t',\,j})^2$, and
    \item $d_{\infty}(\xx_t,\yy_{t'}) = \max_{j=1,\ldots,m} \{|x_{t,\,j}- y_{t',\,j}|\}$.
\end{itemize}
The distance function $d_1$ is often referred to as the \textit{Manhattan} or taxicab distance, $d_2$ as the (squared) \textit{Euclidean} distance, and $d_{\infty}$ as the \textit{Chebyshev} distance.\footnote{In general, the $L^p$ distance between vectors $\xx$ and $\yy$ of length $n$ is defined as \mbox{$||\xx-\yy||_{p} = \sqrt[\uproot{3}p]{(\sum_{i=1}^n |x_i-y_i|^n)}$}. However, we only consider integer-valued input signals in this work and $p$th roots of integers are not always integers for general $p$. This also motivates our choice for the \textit{squared} Euclidean distance instead of the standard Euclidean distance (see also \Cref{sec:conclusions}).}

For two time series $X, Y \in \sX$, we denote the series distance between $X$ and $Y$ by \mbox{$\delta:\sX \times \sX \rightarrow \sZ_+$}. 
A common example of a series distance between two time series of equal length (i.e., $|X| = |Y|= T$) is the sum of the local distances, i.e.,
\begin{equation}
    \sum_{t=1}^T d(\xx_t,\yy_t).
    \label{eq:basic_distance}
\end{equation}
Dynamic Time Warping (DTW) is an alternative to the sum that is more suitable for settings in which time distortions (i.e., delays) are a factor. DTW originates from  speech recognition \cite{berndt1994using}, but its use has become ubiquitous in a vast range of fields from robotics and medicine to cryptanalysis and astronomy \cite{rakthanmanon2012searching}. 

\begin{figure}
\centering
\subfloat[][local distances]{\includegraphics[width=0.325\linewidth]{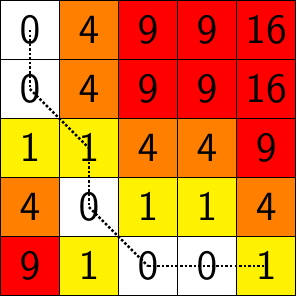}\label{fig:example_1_d_with_path}}
\hspace{0.1\linewidth}
\subfloat[][distance to top-left]{\includegraphics[width=0.325\linewidth]{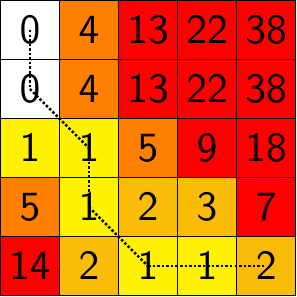}\label{fig:example_1_dd_with_path}}
\caption{Computation of the DTW for two time series of length 5: the dotted line depicts a DTW path.}
\label{fig:dtw_computation}
\end{figure}

Computation of the DTW is equivalent to finding the shortest path in a graph in which the step distances are given by the local distances between the series at different time indices. This is illustrated in \Cref{fig:example_1_d_with_path}, in which each node represents a pair of time series points, and the only valid moves are to the bottom, right, and bottom right. 
The DTW distance between the two series as a whole is given by the shortest path from the top-left to the bottom-right in this graph: this is depicted by the line in  \Cref{fig:example_1_d_with_path}. In \Cref{fig:example_1_dd_with_path}, we depict, for each node, the distance of the shortest path to the top left -- as such, the DTW distance is equal to the value in the bottom right. In contrast, the basic sum is represented by the sum of the values on the diagonal in \Cref{fig:example_1_d_with_path}, which is $4+4+1+1=10$. This is considerably larger than the DTW distance (i.e., 2), which indicates a higher level of similarity than the sum would suggest.

The computation time of DTW distances between time series $\xxx$ and $\yyy$ that respectively consist of $T$ and $T'$ time points is $O(TT')$, i.e., quadratic for equal-length time series. This can be prohibitively expensive for medium-capacity devices such as smartphones.
Furthermore, DTW is not technically a distance \textit{metric} because it does not satisfy the triangle equality, which holds for $\delta$ if for any $X, Y, Z \in \sX$, the following is true: $\delta(X,Y) + \delta(Y,Z) \geq \delta(X,Z)$. The triangle equality implies that a variety of algorithms are be applied, in particular that time series can be \textit{indexed}, which allows for much more efficient nearest-neighbor searches on a set of time series \cite{keogh2005exact,chen2004marriage,marteau2008time}. 
The Time Warp Edit Distance (TWED) is an alternative distance measure designed to capture both temporal alignment and magnitude differences between two time series, and which does satisfy the triangle inequality \cite{marteau2008time}.  
In practice, computation of the TWED is similar to computation of the DTW distance, with the exception that steps to the right and bottom (which represent time shifts) are penalized by a constant $\lambda > 0$ instead of the local distance, plus a factor representing timestamp differences (which we omit as we assume these differences to be negligible).

As we discuss further in \Cref{sec:proof_structure}, all of the methods discussed above can be represented in the same framework. Several other time series methods exists that are similar to the DTW/TWED or sum, and which we do  not consider in detail: these include Keogh's DTW lower bound \ \cite{keogh2005exact,vlachos2006indexing}, edit distance with real penalties \cite{chen2004marriage}, and cosine similarity \cite{steck2024cosine}. The discrete Fréchet distance, which is considered in related work \cite{zhu2014privacy}, is highly similar to DTW with the exception that the max instead of the sum is taken over the local distances in the shortest path.

\subsection{Threat Model}
\label{sec:threat_model}

Our threat model has two types of adversaries. The first is an attacker who seeks to authenticate as the user while not having access to the user's raw biometric data. This attacker has full access to the user's device, e.g., after stealing it, and can log into the \name{} app and the secondary app, and spoof any biometric signal (e.g., blow acoustics) using deepfake techniques. However, the attacker cannot retrieve raw base samples from the device's memory, which is a reasonable assumption if the device supports disk encryption. Furthermore, the attacker is unable to install malicious apps on the device that record the user's input during login attempts prior to gaining access to the device. 
The second threat is the secondary app, which we assume to be honest-but-curious, and which may learn privacy-sensitive data from the proofs sent by the \name{} app. 

We assume that the sensors are not compromised -- i.e., we assume that they produce a faithful representation of the input signal. This is a reasonable assumption when the device uses the sensors on a smartphone with app/component sandboxing, or if an external sensor (e.g., a security camera) validates the data submitted by the \name{} app. We also assume that an attacker who gains access to an external sensor is not able to obtain data involving historical authentication attempts, which may contain information about the user's biometrics. We assume that all communication between the various entities is secure and private, e.g., because it is encrypted via SSL/TLS.

\subsection{Requirements}
\label{sec:requirements}

Our system has the following requirements.
\begin{enumerate}
    \item \textbf{Generality.} The system should be able to support a wide range of time series comparison methods, including the diagonal sum, DTW, and TWED, and  local distance functions, including $d_1$, $d_2$, and $d_{\infty}$.

    \item \textbf{Security.} The probability that an attacker who does not have access to the user's biometric data incorrectly authenticates as the user is smaller than some threshold $p^*$.

    \item \textbf{Privacy.} The honest-but-curious secondary app does not learn information about the user's raw biometric data from the proofs and commitments submitted by \name{}.

    \item \textbf{Local Computations.} Raw biometric data is exclusively processed on the user's device.

    \item \textbf{Efficiency.} Proof generation times, proof verification times, and communication overheads are small enough for practical performance on a smartphone.
\end{enumerate}
We will discuss the first four requirements in more detail in \Cref{sec:analysis}, and the efficiency requirement in \Cref{sec:evaluation}.

\section{The \name{} Protocol}
\label{sec:our_proposal}
\subsection{Workflow}
\label{sec:workflow}

\begin{figure}
\subfloat[][Step 1: Registration]{\framebox{\includegraphics[width=0.45\linewidth]{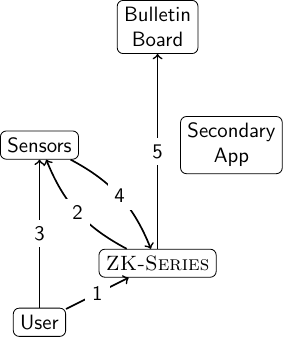}}\label{fig:registration}}
\hspace{0.025\linewidth}
\subfloat[][Step 2: Authentication]{\framebox{\includegraphics[width=0.45\linewidth]{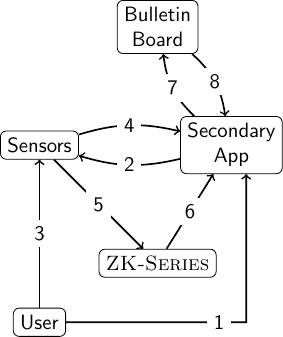}}\label{fig:authentication}}
\caption{\name{} workflow.}
\label{fig:workflow}
\end{figure}

\name{}'s workflow consists of two main phases: registration and authentication. The first phase of the \name{} protocol consists of 5 steps (as depicted in \Cref{fig:registration}) and is executed once, whereas the second phase consists of 8 steps (as depicted in \Cref{fig:authentication}) and is executed whenever the user seeks access to a restricted resource.

\textbf{Registration.}
\begin{enumerate}
\item The user starts the \name{} app and creates an account. 
\item[(2-4)] The \name{} app requests $n$ base readings from the sensors (step 2), which the sensors obtains from the user (step 3). Finally, the sensors send the base readings to the \name{} app (step 4). This can be done actively, i.e., the user is required after registration to record $n$ short samples on the spot, or passively, i.e., after registration, the \name{} periodically queries the sensor for data to gradually build a large database of historical readings.
\setcounter{enumi}{4}
\item The \name{}  app sends commitments of the base readings to the ledger after registration has been completed. Furthermore, the app may periodically post updates to the base samples if the user re-records the base readings, or if readings are passively generated.
\end{enumerate}

\textbf{Authentication.}
\begin{enumerate}
\item The user starts the secondary app, which requires authentication. The secondary app, or the user, then starts the \name{} app to facilitate the authentication process.
\item[(2-5)] Next, the user records $n'$ readings through the sensors. The sensors send the data both to the secondary app and the \name{} app. 
\setcounter{enumi}{5}
\item Next, the \name{} app attempts to generate an authentication proof based on the user's new reading. This proof consists of 1) commitments to the new reading and (some of) the base readings, and 2) a multitude of zero-knowledge proofs involving the values underlying the commitments -- we discuss the latter in more detail in \Cref{sec:proof_structure}. If the proof generation has concluded successfully, then the \name{} app sends the commitments and the zero-knowledge proofs to the secondary app. If not, it asks the user to record another reading (this is not depicted in \Cref{fig:authentication}).
\item[(7-8)] The secondary app verifies that 1) zero-knowledge proofs are valid, 2) the commitments to the new readings are consistent with the data received from the sensors, and 3) the commitments to the base readings are consistent with those on the bulletin board. Upon successful verification, the user is granted access to the resource.
\end{enumerate}

\subsection{Authentication Proof Structure}
\label{sec:proof_structure}

The key technical step of the \name{} protocol is the generation of the authentication proof in step 6 of the authentication phase. The exact composition of these proofs depends on the three main design choices of the \name{} implementation: the local distance function, the series distance function, and the authentication function. We discuss the technical details of these proofs in \Cref{sec:protocols}, and describe their general structure in this section.

\begin{figure}[t]
\includegraphics[width=0.465\linewidth]{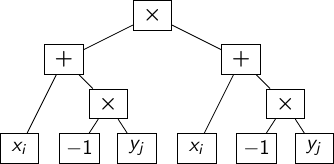}
\hspace{0.4cm}
\includegraphics[width=0.465\linewidth]{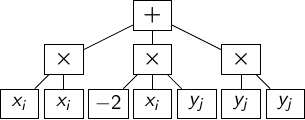}
\label{fig:euclid_tree}
\caption{Two equivalent syntax trees for the computation of $(x_t - y_{t'})^2$, with only the sum operator $+$ and multiplication operator $\times$ in the intermediate nodes and root.}
\end{figure}

In the following, we denote the base series by $(X_1,\ldots,X_n)$  and the new series by $(Y_1,\ldots,Y_{n'})$. For convenience, we say that a function $f:\sZ_p^d \rightarrow \sZ_p$ is \textit{expressible} using a set of operations $\mathcal{O}$ if it can be represented using a syntax tree in which each input value is represented by a unique leaf, intermediate nodes represent operations from $\mathcal{O}$, and the root of the tree contains the output value. For example, two syntax tree representations of the Euclidean distance $(x_y-y_{t'})^2$ are depicted in \Cref{fig:euclid_tree}. For the proofs, we rely on zk-\textit{range proofs}, which assert that an underlying value is within some interval, and zk-proofs of functions expressible as sum, multiplication, and min/max operations. The exact structure of the zk-proofs will be discussed in \Cref{sec:protocols}. As in \cite{zhu2014privacy}, we define a \textit{coupling} $C$ of $X$ and $Y$ as a sequence of pairs of indices $((i_l,j_l))_{l=1,\ldots,k}$ such that $i_l=1,\ldots,|X|$ and $i_l = 1,\ldots,|Y|$, and we define $\mathcal{C}_{X,Y}$ as the set of all valid couplings of $X$ and $Y$.

\textbf{Authentication Proof.}
Given a set of base series $\{X_i\}_{i=1,\ldots,n}$, set of new series $\{Y_j\}_{j=1,\ldots,n'}$, series distance function $\delta$, and function $f_{\delta^*}:\sZ^k \rightarrow \sZ_+$ expressible using sum, min/max, and multiplication operations, the authentication distance is given by
$$
\delta^* = \min_{(i_1,j_1),\ldots,(i_k,j_k)} f_{\delta^*}(\delta(X_{i_1}, Y_{j_1}), \ldots, \delta(X_{i_k}, Y_{j_k})).
$$
That is, the authentication distance $\delta^*$ equals the smallest value of $f_{\delta^*}$ applied to $k$ pairs of from the base series set and the new series set.
The necessary condition for user authentication is then that
\begin{equation}
\delta^* < \theta.
\label{eq:authentication_criterion}
\end{equation}
To prove successful authentication, the 
\name{} app generates:
\begin{enumerate}
\item the indices $(i_1,j_1),\ldots,(i_k,j_k)$,
\item a zk-range proof that $\delta^* < \theta$,
\item a zk-proof that $\delta^*$ has been correctly computed from applying $f_{\delta^*}$ to $\delta(X_{i_1}, Y_{j_1}), \ldots, (X_{i_k}, Y_{j_k})$, and
\item[(*)] zk-proofs for the correct computation of the series distances $\delta(X_{i_1}, Y_{j_1}), \ldots, (X_{i_k}, Y_{j_k})$, as discussed below.
\end{enumerate}
Prominent examples of authentication protocols covered by the above conditions include:

$\bullet$ \textit{Sum} of the distances to the $k$ nearest neighbors.

$\bullet$ \textit{Max} of the distances to the $k$ nearest neighbors.

\textbf{Series Distance Proof.}
Given two time series \mbox{$X,Y \in \sX$}, local distance function $d$, local distance matrix $D = (d(x_i,y_j))_{i=1,\ldots,|X|, j=1,\ldots,|Y|}$, set of acceptable couplings $\mathcal{C}_{X,Y}$, graph edge functions $f_{i,j}$, and total function $f_{\delta}$ expressible using sum, min/max, and multiplication operations, the series distance $\delta(X,Y)$ is given by
$$
\delta(X,Y) = \min_{\mathcal{C}_{X,Y}} f_{\delta}( f_{i_l,j_l} (D)).
$$
To prove correct computation of the series distance, the \name{} app generates:
\begin{enumerate}
\setcounter{enumi}{3}
\item the indices $(i_1,j_1),\ldots,(i_k,j_k)$,
\item zk-proofs of the correct computation of $f_{i_l,j_l}$ over $D$
\item zk-proof of the correct computation of $f_{\delta}$ over $f_{i_l,j_l} (D)$
\item[(*)] zk-proofs of the correct computation of the local distances in $D$, as discussed below.
\end{enumerate}
Prominent examples of series distance functions covered by the above conditions include:

$\bullet$ \textit{Diagonal sum}, with \mbox{$f_{i,j}(D) = d(x_{i}, y_{j})$}, \mbox{$f_\delta(x_1,\ldots,x_k) = \sum_{i=1}^k x_i$}, and \mbox{$\mathcal{C}_{X,Y} = ((1,1)\ldots,(|X|,|X|))$}.

$\bullet$ \textit{DTW}, with $f_{\delta}(x_1,\ldots,x_n) = \sum_{i=1}^n x_i$ 
$$
f_{i,j}(D) = \min \left\{ 
\begin{array}{rl} 
d(x_{i_l-1}, y_{j_l}) & \text{if } i_l > 1,  \\ 
d(x_{i_l}, y_{j_l-1}) & \text{if } j_l > 1, \\ 
d(x_{i_l-1}, y_{j_l-1}) & \text{if } i_l > 1, j_l > 1
\end{array} \right.
$$
and $\mathcal{C}_{X,Y}$ such that $C = ((i_1,j_1),\ldots,(i_k,j_k)) \in \mathcal{C}_{X,Y}$ if and only if $(i_1,j_1) = (1,1)$, $(i_k,j_k) = (|X|,|Y|)$, and for $1<j<k$ $(i_l,j_l) = (i_{l-1}+1, j_{l-1}) \vee (i_{l-1}, j_{l-1}+1) \vee (i_{l-1}+1, j_{l-1}+1)$.

$\bullet$ \textit{Discrete Fr\'echet}, with $f_{\delta}(x_1,\ldots,x_n) = \max_{i=1,\ldots,n} x_i$ and $f_{i,j}$ and $\mathcal{C}_{X,Y}$ as above

$\bullet$ \textit{TWED} with no gap penalty and time shift penalty $\lambda$, with
$$
f_{i,j}(D) = \min \left\{ 
\begin{array}{rl} 
\lambda + d(x_{i_l}, x_{i_l-1}) & \text{if } i_l > 1,  \\ 
\lambda + d(y_{i_l-1}, y_{j_l}) & \text{if } j_l > 1, \\ 
\begin{array}{r} d(x_{i_l}, y_{j_l}) + \\ d(x_{i_l-1}, y_{j_l-1}) \end{array}& \text{if } i_l > 1, j_l > 1
\end{array} \right.
$$
 and $f_\delta$ and $\mathcal{C}_{X,Y}$ the same as for DTW.

\textbf{Local Distance Proof.}
Given two vectors $x,y \in \sZ^m$, and $f_d$ expressible using sum, min/max, and multiplication operations, the local distance function is given by 
$$
d(x,y) = f_d(x_1,\ldots,x_m,y_1,\ldots,y_m).
$$
To prove correct computation of the series distance, the \name{} app generates 
\begin{enumerate}
\setcounter{enumi}{6}
\item a zk-proof of the correct computation of $d(x,y)$ from $f_d$. 
\end{enumerate}
Prominent examples of supported local distance functions include $d_1$, $d_2$, and $d_{\infty}$ as defined in \Cref{sec:data_model}.

\subsection{Analysis}
\label{sec:analysis}

\textbf{Generality.} As discussed in \Cref{sec:proof_structure}, 
\name{} supports all of the protocols mentioned in the generality requirement.

\textbf{Security.} As per \Cref{sec:threat_model}, the main threat to security is an attacker who has gained access to the device running the \name{} app, but not the biometric data. The attacker's goal is to convince the secondary app to accept the proof -- to do so, the attacker must first generate a new reading using the sensors as per steps 2--5 of the authentication phase. Next, the attacker must generate zk-proofs in step 6 that are successfully verified by the secondary app. To succeed, the attacker must do either of the following:
\begin{enumerate}
    \item generate a reading that satisfies \eqref{eq:authentication_criterion}, or
    \item falsely convince the secondary app using an incorrect proof.
\end{enumerate}
The second option only has an infinitesimal probability of success due to the probabilistic correctness of the zk-proofs. Furthermore, the first option is successful with a probability that is upper bounded by the accuracy of the biometric authentication protocol. As such, the system satisfies our notion of security.

We emphasize that an attacker cannot use bisection to get closer to the true biometric data through repeated trials, as the \name{} app does not provide information about the proof structure (including the indices of steps 1 and 5) after unsuccessful attempts.

\textbf{Privacy.} As per \Cref{sec:threat_model}, the main threat to privacy is the honest-but-curious secondary app, which receives the proofs from items 1--8 in \Cref{sec:proof_structure}. The zk-proofs satisfy the property of zero-knowledge, meaning that they reveal nothing about the underlying witnesses. The indices of steps 1 and 5 cannot be used to determine which measurements are closer, since the final distance $\delta^*$ is not revealed as part of the proofs. Finally, unsuccessful proof attempts are not sent to the secondary app, so no information can thus be obtained as well.

\textbf{Local Computations.} All the computations on biometric base readings during the authentication phase are performed in step 6, which is executed entirely on the user device.

\textbf{Efficiency.} This will be demonstrated in \Cref{sec:evaluation}.

\section{Implementation}
\label{sec:implementation}

\subsection{Cryptographic Primitives}
\label{sec:crypto_primitives}

\textbf{Cryptographic Commitments.} 
A cryptographic commitment protocol consists of two functions: the setup function $\textsc{COM.Setup}(1^{\kappa})$ and the commitment function $\textsc{COM.Commit}(P_c,x,r)$. The setup function takes as input a security parameter $\kappa$ and outputs a set of commitment parameters $P_c$. The commitment function takes as input the commitment parameters, a message value $x$, and a hiding factor $r$, and returns a commitment $c$. For brevity, we will use the shorthand notation $\textsc{COM.Commit}(P_c,x,r) = C(x,r)$ and assume that $P_c$ is set globally during initialization (i.e., each \name{} app uses the same commitment parameters regardless of the device).

The commitment function $C$ must satisfy the following properties: 1) it is \textit{hiding}, i.e., it is computationally infeasible for a verifier with knowledge of $c$ to obtain any knowledge of $x$, 2) it is \textit{binding}, i.e., it is computationally infeasible for the prover to find another $x', r'$ such that $C(x,r) = C(x',r')$.
Finally, a cryptographic commitment scheme is said to be \textit{additively homomorphic} if an addition operator on the commitment scheme exists such that
$C_P(x_1,r_1) + C_P(x_2,r_2) = C(x_1+x_2,r_1+r_2)$.

\textbf{Zero-knowledge Proofs.}
Zero-knowledge proofs (ZKPs) enable one party -- the prover -- to convince another party -- the verifier -- that a statement is true without revealing any additional information beyond the validity of the statement itself. 
In particular, let $R_{zk}$ be an efficiently computable relation of the form $(s, w) \in R_{zk}$, where $s$ is a statement and $w$ is a witness of $s$. A non-interactive zero-knowledge proof system for the relation $R_{zk}$ consists of three algorithms. $\textsc{NIZK.Setup}(1^{\kappa'})$ which takes as input a security parameter $\kappa'$ and outputs system parameters $P_{zk}$. $\textsc{NIZK.Prove}(P_{zk},s,w)$ takes as input the system parameters $P_{zk}$ and a statement-witness pair $(s, w)$ and outputs a proof $\pi$. $\textsc{NIZK.Verify}(P_{zk},s,\pi)$ takes as input $P_{zk}$, a statement $s$, and a proof $\pi$, and outputs $1$ (true) or $0$ (false). The proof system NIZK is \textit{zero-knowledge} if the generated proofs reveal nothing about the witnesses, and has \textit{simulation-extractability} if for any proof generated by the adversary, there exists an efficient algorithm to extract the corresponding witnesses. 

In \name{}, we use two different NIZK protocols: zk-multiplication proofs (ZKMPs), which prove relations of the form
\begin{equation*}
\begin{split}
R_{\text{ZKMP}} & = \{(c_{x}, c_{y}, c_{xy}), (x, y, r_x, r_y, r_{xy}) \\ & \;\;|\;c_x = C(x, r_x), c_y = C(y, r_y)  \text{ and } c_{xy} = C(xy, r_{xy})\}
\end{split}
\end{equation*}
and zk-range proofs (ZKRPs) for relations of the form
$$
R_{\text{ZKRP}} = \{(c, v^*), (v, r)\;|\;c = C(v, r) \text{ and } v \in [0, v^*)\}.
$$
For ZKMPs, we use the protocol by Fujisaki and Okamoto \cite{fujisaki1997statistical}, which is also used by BioZero. For ZKRPs, multiple approaches exist, as we discuss in more detail below.

\textbf{Zero-knowledge Range Proofs.}
State-of-the-art ZKRP protocols can be divided into four main categories \cite{christ2024sok}. The first category uses generic proof systems for arithmetic circuits to express the range proof's relation. zk-SNARKs \cite{bitansky2012extractable,groth2016size} and zk-STARKs \cite{ben2018scalable} are prominent examples of zero-knowledge proof systems for arithmetic circuits. 
The second category uses $n$-ary decomposition \cite{bunz2018bulletproofs,wang2024swiftrange,eagen2024bulletproofs}, in which a value $x$ is decomposed into a sum $\sum_{i=1}^{l-1} c_i n^i$ with $c_i \in \{0,1\}$ to prove that $x \in [0, n^l)$. Bulletproofs \cite{bunz2018bulletproofs}, which use binary decomposition, are a prominent example of such a protocol; due to their high efficiency and small proof sizes, they have grown to underpin a wide variety of real-world applications, including the Monero cryptocurrency. 
The third category uses square decomposition, and a recently proposed method, \Sharp{} range proofs \cite{couteau2022sharp}, belong to this category. 
The fourth and final category is that of hash-based range proofs \cite{chalkias2021hashwires} -- this technique works only for a specific type of commitment schemes (i.e., hash-based commitments) that is beyond the scope of this work.

Previous work \cite{christ2024sok} has found that the computation times of Groth16 (which is used by BioZero to prove that the total distance is smaller than the threshold $\theta$) for both prover and verifier are an order of magnitude worse than for ZKRPs of any of the other categories. Meanwhile, although $n$-ary decomposition techniques such as Bulletproofs were found to have short proof sizes and verification times, this came at the cost of long proof generation times. This is appropriate for settings with a large discrepancy between the computational resources of the prover and verifier, e.g., a proof generated by a high-end machine but verified on a public blockchain. However, in our setting, the prover runs on a low-capacity device such as a smartphone, so short proof generation times are critical. By contrast, \Sharp{} was found to achieve considerably shorter proof generation times than the other categories, and is therefore best suited for our setting. We discuss \Sharp{} in more detail below.

\textbf{\Sharp{} Range Proofs.}
In their most concise form,\footnote{Alternatively, \Sharp{} range proofs can be derived from Lagrange's four-squares theorem, but this leads to larger proof sizes.} \Sharp{} range proofs rely on Legendre's three-square theorem, which (as a corollary) states that for any integer $x$, there exist integers $y_1,y_2,y_3 \geq 0$ such that \mbox{$4x+1 = y_1^2 + y_2^2 + y_3^2$} if and only if $x \geq 0$. This can be used to prove statements of the form $x \in [0, B]$ for some integer $B$ by providing $y_1,y_2,y_3 \geq 0$ such that $4x(B-x)+1 = y_1^2 + y_2^2 + y_3^2$. 
Although we defer to \cite{couteau2022sharp} for the technical details, the core innovation of \Sharp{} is a non-trivial proof of `shortness' of the variables $y_1,y_2,y_3$, which is probabilistic in the sense that for given $R>0$, the security property holds with probability $2^{-R}$. From a practical perspective, the values $x$ and $B$ must also be short because decomposition of $4x(B-x)+1$ into three squares can be computationally expensive for moderately large integers, as we discuss in more detail below. A key property of \Sharp{} ZKRP is that they support efficient batching of multiple range proofs: although execution times asymptotically depend on $R$ and $n$, in practice the proof generation times are dominated by operations on commitments, which in \Sharp{} do not depend on $R$ and $n$. Finally, the required shortness of the encrypted values in \Sharp{} is not an obstacle in our setting as we have found that rounding and rescaling all values to ensure they are between $0$ and $1\,000\,000$ has a negligible impact on accuracy (we omit further discussion of this due to space limitations).

\subsection{Range Proof Adaptation}
\label{sec:protocols}

\textbf{Proof Structure.}
An authentication proof, as defined in \Cref{sec:proof_structure}, consists of 7 main components. Of these, two are index pair sequences (items 1 and 4), one is a ZKRP that proves that the authentication distance is within an acceptable threshold (item 2), and four are zk-proofs of the correct computation of a function expressible using um, min/max/ and multiplication operations (items 3, 5, 6, and 7). Proofs of the last category are structured as follows: leaf nodes are included in the proof as commitments (if they correspond to biometric data) or as scalar values. For all intermediate nodes and the root node, a commitment to the value in the node is included. For sum nodes, verification is trivial because of the additive homomorphism of our commitment scheme. For multiplication nodes, a ZKMP is provided. For min/max nodes with two child nodes, a ZKRP is given that $x>y$ if $max(x,y) = x$ or $min(x,y) = y$, and for $y>x$ otherwise. Finally, min/max and multiplication nodes with more than two children can be decomposed into repeated 2-child nodes because $x \times y \times z=x \times (y\times z)$ and $\max(x,y,z) = \max(x, \max(y,z))$.
A key insight for efficiency that all ZKRPs in the above can be aggregated into a single \Sharp{} proof, which significantly reduces computation times. We do note that Bulletproofs also support proof batching, but that this only affects the resulting proof size and not the generation times.

\textbf{Efficient Decomposition.}
As mentioned previously, we aggregate the ZKRPs for each min/max intermediate node and the final authentication proof into a single \Sharp{} ZKRP. However, the value $x$ underlying each individual proof must still be decomposed into three squares. This can be done efficiently in the following way \cite{lu2024efficient}: if we assume that each sufficiently large integer can be decomposed into the sum of a square and a prime $p'$ such that $p' = 1 \text{ mod } 4$ \cite{rabin1986randomized}, then we can pre-compute for all such primes their two-square decomposition (which must exists by Fermat's theorem on sums of two squares). For any integer, we then iterate over the cached primes and check whether $x - p'$ is a square -- if so, $\sqrt{z-p'}$ and the precomputed 2-square decomposition of $p'$ are a valid 3-square decomposition of $x$. Storing only the primes is efficient because the frequency of primes among larger integers decreases logarithmically: e.g., all 2-square decompositions for $p' < 1\,000\,000$ can be stored in a 600KB CSV file.

\subsection{Software Implementation}
\label{sec:software_implementation}

The \name{} app as discussed in \Cref{sec:our_proposal} has the following functionality: a frontend that allows users to register and record readings, and a backend that implements the proof generation as discussed in \Cref{sec:proof_structure}. We focus on the backend, as this is most technically complex part of the protocol. 
Proof generation consists of two main steps: 1) the time series analysis protocols, particularly DTW and TWED, and 2) the cryptographic protocols, particularly the ZKMPs and ZKRPs.
The first component is relatively straightforward, as are ZKMPs: similar to BioZero \cite{lai2024biozero}, we have implemented ZKMPs using \cite{fujisaki1997statistical}. Construction of a single ZKMP consists of computing 6 commitments via 9 (big) integers, and verification of 4 commitment additions and 5 scalar multiplications, which is relatively inexpensive. 

Instead, the most challenging part of the implementation is the implementation of the ZKRPs, particularly because no publicly available implementation of \Sharp{} exists. To illustrate the performance advantages of \Sharp{}, we have also implemented the ZKRPs using Bulletproofs. Because of Android's built-in support for Kotlin, we chose the Bulletproofs implementation in Java \cite{weavechainbp} by weavechain as a baseline for the Bulletproof ZKRPs. This library is in turn based on the Rust implementation of Bulletproofs \cite{zkcryptobp}, which uses Pedersen commitment using the curve25519 elliptic curve, and which supports curve compression using Ristretto (which is an extension of Decaf \cite{hamburg2015decaf}). We implemented \Sharp{} ZKRPs on the same elliptic curve library as weavechain's Bulletproofs implementation to allow for a fair comparison in \Cref{sec:scalability}.

\section{Experimental Evaluation}
\label{sec:evaluation}
In this section, we present an empirical evaluation of our \name{} implementation. 
The goal of our evaluation is twofold: first, to motivate the use of a wider range of time series distance functions than the Euclidean sum as implemented in BioZero. We perform experiment for two use cases, shake-based and blow-based authentication, and find that the Manhattan distance has superior performance to the Euclidean distance, and that DTW has superior performance to the sum for shake gestures. The second goal is to show that our \name{} implementation satisfies the efficiency requirement of \Cref{sec:requirements}, i.e., that it has practical performance on a mid-range smartphone. 

We conducted our experiments on a Galaxy A71 smartphone, a model that was launched in November 2019. It has an Octa-core CPU (2x2.2 GHz Kryo 470 Gold \& 6x1.8 GHz Kryo 470 Silver), 128GB of memory, and 8GB RAM. Its operating system is Android v13.

\subsection{Shake-Based Authentication}

Our first case study leverages the motion sensors in a typical Android smartphone, i.e., the accelerometer, the gyroscope, magnetometer, and orientation sensor. Each sensor records three coordinates with a frequency of 5--50 Hz, and the measurements at each time slot can therefore be represented as an array of $m=12$ elements. Upon registration, the user generates $n$ base readings that each consists of the user shaking the phone for a number of seconds. Recording starts when the phone registers significant movement, and stops when it no longer detects movement or after a fixed maximum duration. The aim is for the shake patterns to be similar -- after recording $n$ samples, the users can be asked to re-record  readings to remove outliers and improve accuracy.
To authenticate, the user shakes a single time and the new reading is compared to the base measurements. We choose the following approach: we identify the $k$ base readings that are closest to the new measurement. If the sum of the distances between the new reading and the $k$ nearest readings is below some threshold, then the user is successfully authenticated. 

To create the dataset, a group of 20 participants (15 male and 5 female) were asked to a gesture consisting of 3 shake motions. They were asked to repeat this 10 times, resulting in 10 readings per participant. For each reading, 12 different input signals were measured: the $x$, $y$, and $z$ coordinates of the accelerometer, gyroscope, and magnetometer, and the yaw, pitch and roll coordinates of the orientation sensor. For illustrative purposes, the acceleration readings of participants 1 and 2 are depicted in Figure \ref{fig_participant1_acc} and \ref{fig_participant2_acc} in the appendix, respectively. The sampling frequency was set to 50 Hz, and the app was programmed to stop recording as soon as the participant finished the third shake. As a result, different time series may have different length: the average, median, minimum, and maximum lengths were $52.91$, $50$, $42$, and $97$ elements, respectively. We note that some measurements are negative -- however, each measurement $x_{t,j}$ was normalized as $K \cdot \frac{x_{t,j} - \min_j}{\max_j - \min_j}$ where $K=1\,000\,000$ and $\min_j$ and $\max_j$ denote the minimum and maximum value for input signal $j \in 1,\ldots,m$, respectively. For $K$, we note that there is a tradeoff between setting it too low (which would reduce accuracy) and too high (which requires a larger list of precomputed 2-square decompositions).
The dataset used for our experiments can be found online \cite{zkseries_data}.

\begin{table*}[]
\centering
    \caption{Accuracy, FPR, precision, and recall for shake-based authentication with $k=1$ and different target recall values ($q=10$, $q=9$, and $q=8$). The highest accuracy and lowest FPR values for each $q$ are highlighted in {bold}.}
    \label{tab:shakeauth_k1}
    \begin{tabular}{cc|cccc|cccc|cccc}
    \multicolumn{2}{c}{distance} & \multicolumn{4}{c|}{$q=10$} & \multicolumn{4}{c}{$q=9$} & \multicolumn{4}{c}{$q=8$}\\[0.1cm]
    series & local & acc. & FPR & prec. & rec. & acc. & FPR & prec. & rec. & acc. & FPR & prec. & rec. \\ \toprule

         & $d_{1}$       & $0.8153$ & $0.1945$ & $0.21$ & $1.00$ & $0.9605$ & $0.0363$ & $0.57$ & $0.90$ & $0.9820$ & $0.0087$ & $0.83$ & $0.80$ \\
        sum & $d_{2}$    & $0.4193$ & $0.6113$ & $0.08$ & $1.00$ & $0.5673$ & $0.4505$ & $0.10$ & $0.91$ & $0.5850$ & $0.4276$ & $0.09$ & $0.83$ \\
         & $d_{\infty}$  & $0.8210$ & $0.1884$ & $0.22$ & $1.00$ & $0.9610$ & $0.0361$ & $0.54$ & $0.91$ & $0.9813$ & $0.0092$ & $0.68$ & $0.80$ \\  \midrule
         & $d_{1}$       & $\bf 0.9065$ & $\bf 0.0984$ & $0.35$ & $1.00$ & $\bf 0.9912$ & $\bf 0.0042$ & $0.92$ & $0.91$ & $\bf 0.9895$ & $\bf 0.0008$ & $0.98$ & $0.80$ \\
        DTW & $d_{2}$    & $0.4830$ & $0.5442$ & $0.09$ & $1.00$ & $0.4967$ & $0.5250$ & $0.08$ & $0.91$ & $0.4977$ & $0.5195$ & $0.08$ & $0.83$ \\
         & $d_{\infty}$  & $0.8938$ & $0.1118$ & $0.32$ & $1.00$ & $0.9895$ & $0.0058$ & $0.81$ & $0.90$ & $0.9892$ & $0.0013$ & $0.79$ & $0.81$ \\  \midrule
         & $d_{1}$       & $0.7473$ & $0.2661$ & $0.17$ & $1.00$ & $0.8867$ & $0.1147$ & $0.30$ & $0.92$ & $0.9410$ & $0.0518$ & $0.45$ & $0.80$ \\
        TWED & $d_{2}$   & $0.8022$ & $0.2082$ & $0.20$ & $1.00$ & $0.9198$ & $0.0795$ & $0.37$ & $0.91$ & $0.9430$ & $0.0503$ & $0.46$ & $0.82$ \\
         & $d_{\infty}$  & $0.7463$ & $0.2671$ & $0.16$ & $1.00$ & $0.8893$ & $0.1116$ & $0.29$ & $0.91$ & $0.9308$ & $0.0626$ & $0.37$ & $0.80$ \\  \bottomrule
         
    \end{tabular}
    
\end{table*}

To evaluate the accuracy of our authentication protocol we perform the following experiment. For authentication, we use $k$-NN max: for given $k$, we determine the distance of a new reading to the $k$ nearest base readings, and if the largest among these (i.e., the $k$th nearest) reading has a distance smaller than the threshold $\theta$, then we authenticate the user. 
For each participant, we treat the 10 readings as their complete set of base readings. To determine an appropriate threshold $\theta$, we use the following procedure: we determine for each reading the distance to each of the other 9 readings. For given $q \in \{1,\ldots,10\}$, we then set the threshold such that $q$ of the user's readings would lead to successful authentication. Although this leads to a target recall of $q/n$, the observed recall may be higher: e.g., a threshold set to accept at least 9 readings may occasionally also accept the 10th reading if the authentication distance is exactly the same. This happens with non-negligible frequency due to symmetries in the dataset: the distance between reading $i$ and $j$ for each user is the same as between reading $j$ and $i$. To compute the ``sum'' series distance for two time series with unequal length, we take the sum over the largest number of elements that they have in common.

There is a notable tradeoff that must be considered when choosing $q$: for $q=10$, all of the user's base readings would lead to authentication, but the threshold may be too permissive when other users try to authenticate. To evaluate this tradeoff, we performed a second experiment where for each user, we treat the $190$ readings by other users as incorrect log-in attempts. Each time a user's reading leads to incorrect authentication (i.e., a false positive) is treated as a security fault. The results of this experiment for $k=1$ and the $3$-dimensional orientation data are depicted in \Cref{tab:shakeauth_k1} for $q=8,9,10$. Each table contains for all combinations of the local distance functions $d_1$, $d_2$, and $d_{\infty}$, and for the series distance functions ``sum'', DTW, and TWED, the accuracy (acc.; fraction correct among total of 4000 attempts), false positive rate (FPR; fraction of false positives among 3800 incorrect attempts), precision (prec.; fraction of true positives among the true and false positives),  and the recall (rec.; fraction of true positives among 200 correct attempts),

From Figure~\ref{tab:shakeauth_k1} we clearly observe the tradeoff between precision and recall: for $q=10$, no user reading results in a false negative, but the fraction of false positives is considerable. The best performing approach is DTW combined with $d_1$ for $k=1$, but even this approach has a false positive fraction of $9.84\%$. This is far too high for a practical authentication protocol. In contrast, for $q=9$ and $q=8$ we observe that if we set the threshold lower, i.e., to reject at least from one each user's base readings, that the fraction of false positives also decreases. In particular, the best performing approach (DTW combined with $d_1$, for $k=1$) has a false positive fraction of $0.08\%$. This is more than two orders of magnitude lower than for $q=10$, and therefore better suited for authentication. We do note that too many false negatives will lead to a diminished user experience. For the TWED, we used a constant $\lambda = 1\,000\,000$: although better choices for $\lambda$ may exist for this particular case study, finding such a value is beyond the scope of this paper.

\subsection{Blow-Based Authentication}\label{sec:blow_based_auth_evaluation}

\begin{table*}[]
\centering
    \caption{Accuracy, FPR, precision, and recall for blow-based authentication with $k=1$ and different target recall values ($q=10$, $q=9$, and $q=8$). The highest accuracy and recall values for each choice of $q$ are highlighted in {bold}.}
    \label{tab:blowauth}
    \begin{tabular}{cc|cccc|cccc|cccc}
    \multicolumn{2}{c}{distance} & \multicolumn{4}{c|}{$q=10$} & \multicolumn{4}{c}{$q=9$} & \multicolumn{4}{c}{$q=8$}\\[0.1cm]
    series & local & acc. & FPR & prec. & rec. & acc. & FPR & prec. & rec. & acc. & FPR & prec. & rec. \\ \toprule
             & $d_{1}$       & $\bf 0.9502$ & $\bf 0.0508$ & $0.29$ & $1.00$ & $\bf 0.9864$ & $\bf 0.0120$ & $0.61$ & $0.91$ & $\bf 0.9907$ & $\bf 0.0058$ & $0.74$ & $0.82$ \\
        sum & $d_{2}$    & $0.9125$ & $0.0893$ & $0.19$ & $1.00$ & $0.9792$ & $0.0193$ & $0.49$ & $0.91$ & $0.9889$ & $0.0075$ & $0.69$ & $0.81$ \\
         & $d_{\infty}$  & $\bf 0.9502$ & $\bf 0.0508$ & $0.29$ & $1.00$ & $\bf 0.9864$ & $\bf 0.0120$ & $0.57$ & $0.91$ & $\bf 0.9907$ & $\bf 0.0058$ & $0.64$ & $0.82$ \\  \midrule
         & $d_{1}$       & $0.9335$ & $0.0679$ & $0.23$ & $1.00$ & $0.9710$ & $0.0276$ & $0.40$ & $0.90$ & $0.9828$ & $0.0136$ & $0.55$ & $0.81$ \\
        DTW & $d_{2}$    & $0.9240$ & $0.0776$ & $0.21$ & $1.00$ & $0.9586$ & $0.0404$ & $0.31$ & $0.91$ & $0.9719$ & $0.0247$ & $0.40$ & $0.80$ \\
         & $d_{\infty}$  & $0.9335$ & $0.0679$ & $0.23$ & $1.00$ & $0.9710$ & $0.0276$ & $0.38$ & $0.90$ & $0.9828$ & $0.0136$ & $0.48$ & $0.81$ \\  \midrule
         & $d_{1}$       & $0.8351$ & $0.1683$ & $0.11$ & $1.00$ & $0.9042$ & $0.0958$ & $0.16$ & $0.90$ & $0.9385$ & $0.0589$ & $0.22$ & $0.81$ \\
        TWED & $d_{2}$   & $0.8150$ & $0.1888$ & $0.10$ & $1.00$ & $0.8837$ & $0.1167$ & $0.14$ & $0.90$ & $0.9325$ & $0.0649$ & $0.20$ & $0.81$ \\
         & $d_{\infty}$  & $0.8351$ & $0.1683$ & $0.11$ & $1.00$ & $0.9042$ & $0.0958$ & $0.16$ & $0.90$ & $0.9385$ & $0.0589$ & $0.21$ & $0.81$ \\  \bottomrule
    \end{tabular}
\end{table*}

Our second case study examines a blowing-based behavioral biometric technique designed for smartphone user authentication. The underlying premise is that the manner in which users blow on a phone screen produces distinctive acoustic patterns, which can serve as a unique behavioral biometric identifier for effective user identification or authentication.
To validate this intuition, a proof-of-concept application was implemented on the Android platform to capture the users' blowing acoustic signals as time series, specifically focusing on audio amplitude data. During data collection, users were instructed to blow on the phone screen in any manner they preferred for a duration of five seconds. The procedure was repeated for 10 sessions 
for each user. Consequently, an experimental dataset was compiled from 50 participants and has been made available via \cite{zkseries_data}.  
For illustrative purposes, the blow acoustic time series data of four participants is depicted in Figure \ref{fig_sample_blow_acoustic_data} in the appendix while the whole dataset can  be found online \cite{zkseries_data}.

We have found that applying a simple moving average filter to the data improves authentication accuracy: in particular, given step size $k$ and window size $w$, 
the smoothed time series $\tilde{\xx}$ of a series $\xx$ with size $T$ is given as follows: let $\xx^i = (\xx_{ik},\xx_{ik+1},\ldots,\xx_{\min(ik+w,T)})$. 
Then $\tilde{x}_i = \frac{1}{|\xx^i|} \sum_{j=1}^{|\xx^i|} \xx^i_j$ for $i=1,\ldots,\lceil T/k \rceil$. This has the added advantage of reducing the size of the time series from $T$ to $\lceil T/k \rceil$, which considerably reduces DTW and TWED computation times even for moderate chocies of $k$. For our experiments, we chose $k=4$ and $w=8$, which reduces the size of our readings from around 250 to 62 elements.

The results for blow-based authentication are displayed for $k=1$ in \Cref{tab:blowauth}. In contrast to the shake-based results, the sum outperforms DTW and TWED for $q=8$. The reason is that the timing behavior in the blow pattern, e.g., waiting for a second before commencing, may be useful to distinguish between users -- in the sum, such differences are reflected, whereas the time shift insensitivity of the DTW and TWED lessens such effects. We have found that the use of a Sakoe-Chiba band \cite{keogh2005exact} improves DTW's performance considerably, but we omit this due to space limitations. (This can be achieved by modifying the set of acceptable couplings $\mathcal{C}_{X,Y}$ for DTW in \Cref{sec:proof_structure}.) We observe that $d_1$ and $d_{\infty}$ always produce the same result because $m=1$. We also observe that the squared Euclidean distance, $d_2$, again consistently has the worst performance, which motivates the support for techniques beyond the Euclidean sum.

Overall, the best observed FPR for blow-based authentication is lower than for shake-based authentication: 0.58\% for $d_1$ and the sum for the former, compared to 0.08\% for $d_1$ and DTW for the latter. However, in both cases we imagine the behavioral biometric to be one part of a multi-factor procedure: e.g., in combination with each other or with a hard biometric such as face recognition. We leave this to further research.

\subsection{Scalability}
\label{sec:scalability}

To investigate the impact of the time series size on computation times, we repeatedly generate synthetic time series data and calculate the average time cost for generating and validating the authentication proof. To generate synthetic data, we use a random number generator to generate $n$ time series with parameters $m$ (the dimension of each time series element) and $T$ (the length of the time series). 

The result for $m=3$ and varying choices of $T$, the time cost of the various protocol components  are displayed in \Cref{tab:comp_times}. We first observe that the time cost of Bulletproofs increases linearly for increasing $T$, which is to be expected as batching does not improve proof generation times. We do observe a mild linear increase in the generation times for \Sharp{} proofs: from $0.972$s for $T=50$ to  $1.120$s for $T=300$. As mentioned in \Cref{sec:implementation}, the reason is that \Sharp{} proof generation times are dominated by operations on commitments, i.e., on elliptic curves, which are independent of $T$. Verification times are approximately $10 \times$ and $60 \times$ lower for Bulletproofs and \Sharp{} proofs, respectively -- for \Sharp{}, the verification times are so small that random noise is greater than any observable increase in computation times due to $T$. The total verification times for \Sharp{} support practical efficiency: e.g., for the best-performing shake-based case study we have $T \in \{42,\ldots, 97\}$, $m=3$, $k=1$, and for $3 * 97 \approx 300$ we have a total verification time of $1.120 + 0.018 = 1.1138$ seconds, which is acceptable from the perspective of the user.

For the DTW calculation, we observe quadratically increasing computation times as discussed in \Cref{sec:data_model}. However, $m$ and $k$ only have a linear impact on the DTW. As such, the impact of this part of the total time to execute \name{} for the shake-based case study as mentioned above is around 3 times $0.056$ (the value for $T=100$), increasing the total authentication time from $1.1138$ to $1.2818$ seconds, which remains acceptable.

\setlength{\tabcolsep}{3.5pt}
\begin{table}[b]
    \centering
    \caption{Computation times (in seconds) for ZKRP generation and verification times for Bulletproofs and \Sharp{}, DTW, and diagonal sum calculations, for varying values of $T$ (the number of time series elements).}
    \label{tab:comp_times}
    \begin{tabular}{l|cccccc}
         & \multicolumn{6}{c}{time (s)} \\ 
         \multicolumn{1}{c|}{$T$} & 50 & 100 & 150 & 200 & 250 & 300 \\ \toprule
         Bulletproofs (gen) & 18.27 & 34.97 & 53.69 & 70.27 & 87.62 & 106.0 \\
         Bulletproofs (vrf) & 2.143 & 3.674 & 5.918 & 7.356 & 9.302 & 11.20 \\
         \Sharp{} (gen) & 0.972 & 0.983 & 1.010 & 1.057 & 1.090 & 1.120 \\
         \Sharp{} (vrf) & 0.022 & 0.017 & 0.016 & 0.016 & 0.018 & 0.018  \\ \midrule
         diag sum & 0.001 & 0.001 & 0.001 & 0.004 & 0.003 & 0.006 \\
         DTW & 0.099 & 0.056 & 0.199 & 0.479 & 0.919 & 1.636 \\ \bottomrule
    \end{tabular}
\end{table}
\setlength{\tabcolsep}{5pt}

\section{Conclusion}
\label{sec:conclusions}
We have presented \name{}, an authentication protocol that provides privacy and efficiency to a wide range of time series comparison protocols. We have provided an implementation of \name{} for three different local distance functions, four different time series distance functions, and two authentication protocols beyond a 1-to-1 comparison. We have presented experimental results that demonstrate that the time series techniques supported by \name{} have higher accuracy than those provided by BioZero \cite{lai2024biozero}, and that, unlike \cite{zhu2014privacy,liu2019privacy,zheng2021efficient}, \name{} has practical performance on a smartphone.

Regarding future work, one interesting direction is to also include the true Euclidean local distance, i.e., the square root of $d_2$ as defined in \Cref{sec:data_model}.
To prove the correct computation of $x = \lfloor\sqrt{y}\rfloor$, we can provide two ZKRPs to show that $x^2 \leq y$ and $(x+1)^2 > y$. Such ZKRPs can be efficiently batched with the others. Another direction is to investigate the loss of privacy from revealing the time index pairs in steps (1) and (4) of \Cref{sec:proof_structure}.
A third direction is to extend \name{} to (some of) the other time series distance functions presented, e.g., the shapelet-based methods discussed in \cite{ruiz2021great} or neural networks. A fourth direction is to apply \name{} to other case studies, e.g.,  facial gestures converted into time series using feature extraction, or continuous authentication in which vast amounts of base readings are generated passively.

\section*{Acknowledgments}
This research is supported by the National Research Foundation, Singapore and Infocomm Media Development Authority under its Trust Tech Funding Initiative (DTC-T2FI-CFP-0002). Any opinions, findings and conclusions or recommendations expressed in this material are those of the author(s) and do not reflect the views of National Research Foundation, Singapore and Infocomm Media Development Authority.

\bibliographystyle{plain}
\bibliography{ref}

\section*{Ethics Considerations}
The data for the shake-based and blow-based authentication case studies were gathered from human subjects. The datasets have been anonymized, and do not contain personally identifiable biometric information: the data collection period for shake gestures is too short to reveal personally identifiable information (e.g., gait), and the frequency for the blow acoustics (50 Hz) is too low to obtain characteristic voice information (which typically requires between 300 and 4000 Hz). As such, we believe that participation in our study constitutes minimal risk.

\appendix
\counterwithin{figure}{section}
\counterwithin{table}{section}
\counterwithin{equation}{section}
\section{Appendix}
\label{sec:appendix}
\begin{figure*}
     \centering
     \begin{subfigure}{0.49\textwidth}
         \centering
         \includegraphics[width=\linewidth]{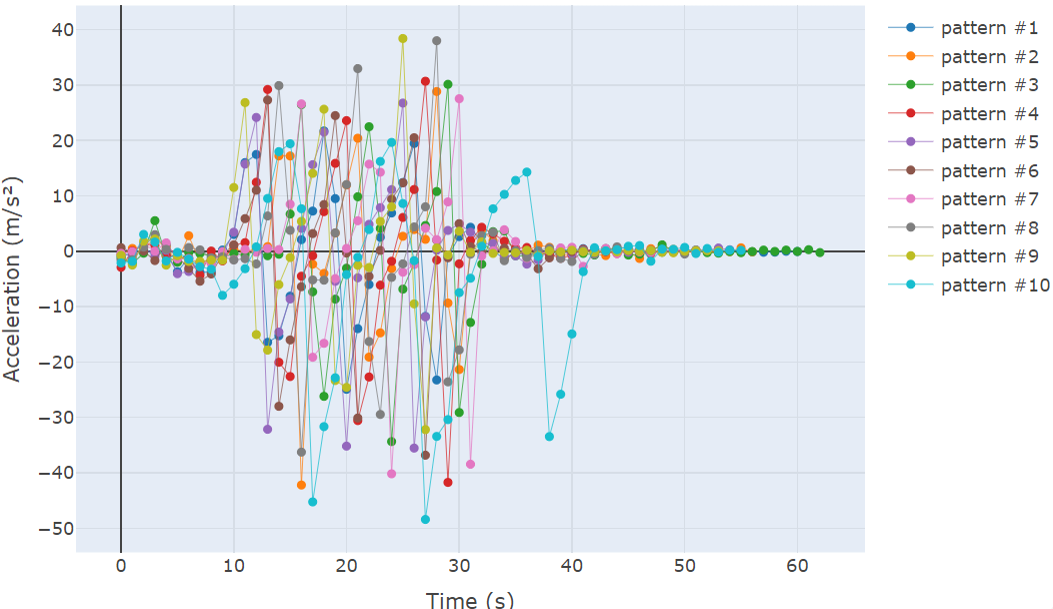}
         \caption{}
         \label{fig_participant1_accx}
     \end{subfigure}
     \begin{subfigure}{0.49\textwidth}
         \centering
         \includegraphics[width=\linewidth]{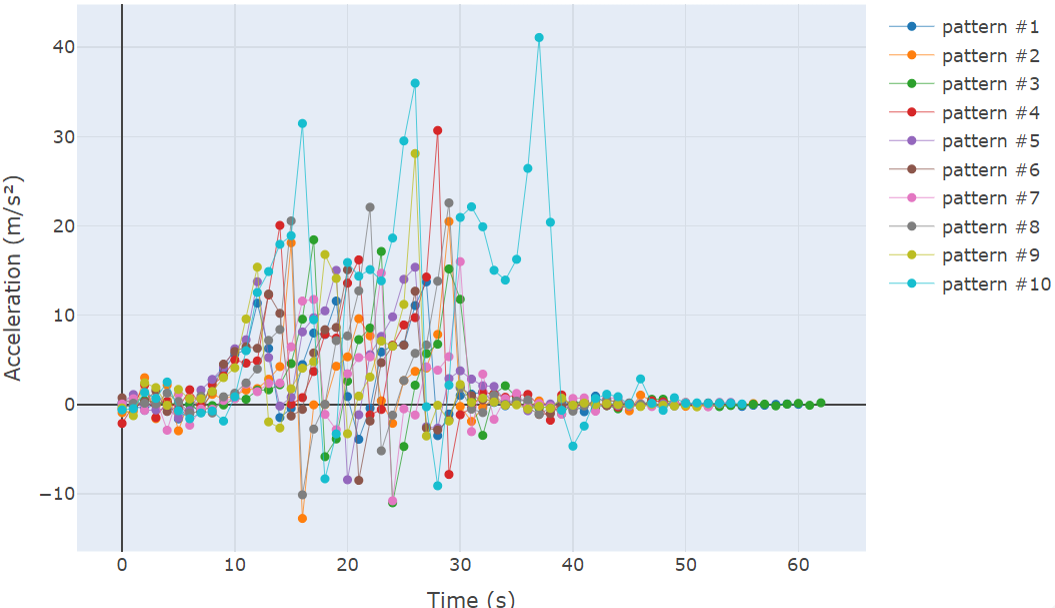}
         \caption{}
         \label{fig_participant1_accy}
     \end{subfigure}
      \begin{subfigure}{0.49\textwidth}
         \centering
         \includegraphics[width=\linewidth]{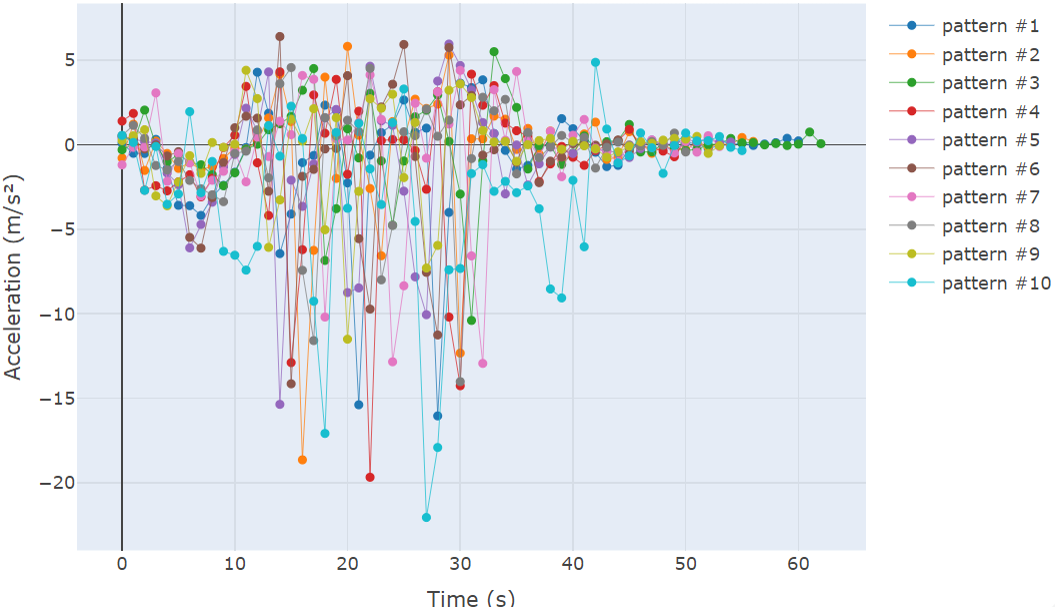}
         \caption{}
         \label{fig_participant1_accz}
     \end{subfigure}
      \begin{subfigure}{0.49\textwidth}
         \centering
         \includegraphics[width=\linewidth]{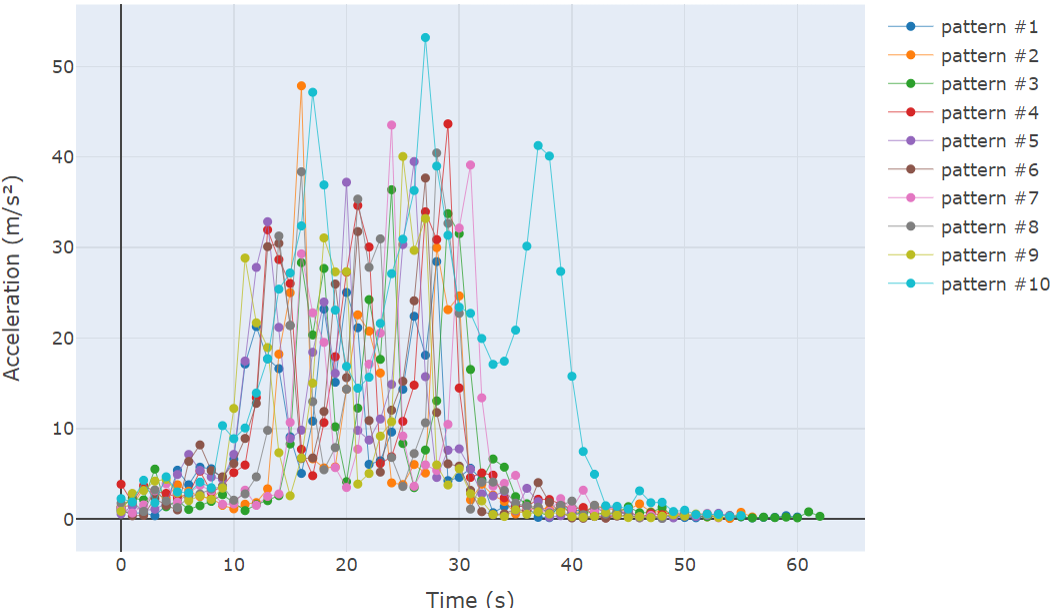}
         \caption{}
         \label{fig_participant1_accR}
     \end{subfigure}
     \caption{Acceleration readings of Participant 1 in: (a) x-axis (b) y-axis (c) z-axis (d) resultant vector (an aggregated data of the readings along the three axes)}
     \label{fig_participant1_acc}
\end{figure*}

\begin{figure*}
     \centering
     \begin{subfigure}{0.49\textwidth}
         \centering
         \includegraphics[width=\linewidth]{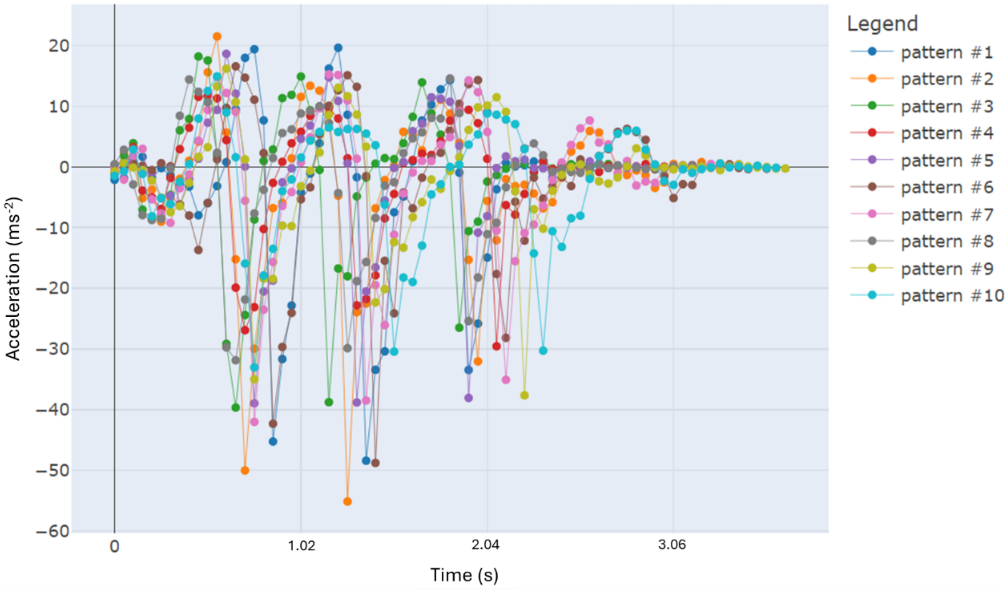}
         \caption{}
         \label{fig_participant2_accx}
     \end{subfigure}
     \begin{subfigure}{0.49\textwidth}
         \centering
         \includegraphics[width=\linewidth]{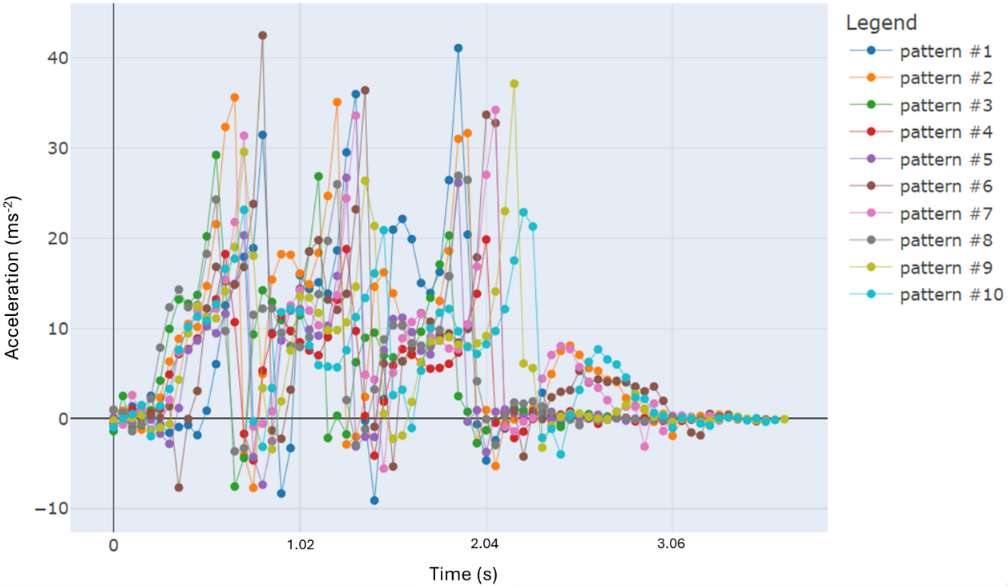}
         \caption{}
         \label{fig_participant2_accy}
     \end{subfigure}
      \begin{subfigure}{0.49\textwidth}
         \centering
         \includegraphics[width=\linewidth]{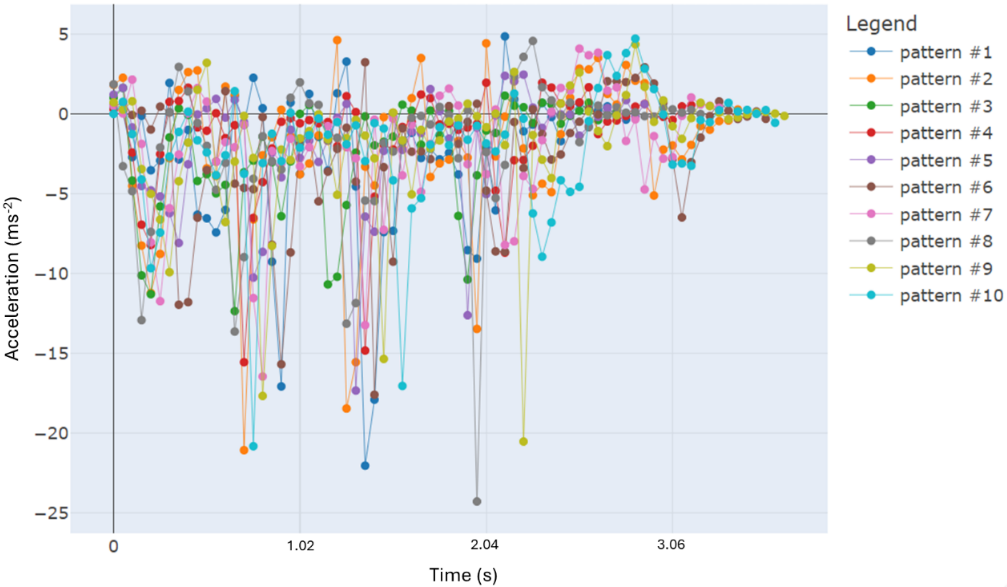}
         \caption{}
         \label{fig_participant2_accz}
     \end{subfigure}
      \begin{subfigure}{0.49\textwidth}
         \centering
         \includegraphics[width=\linewidth]{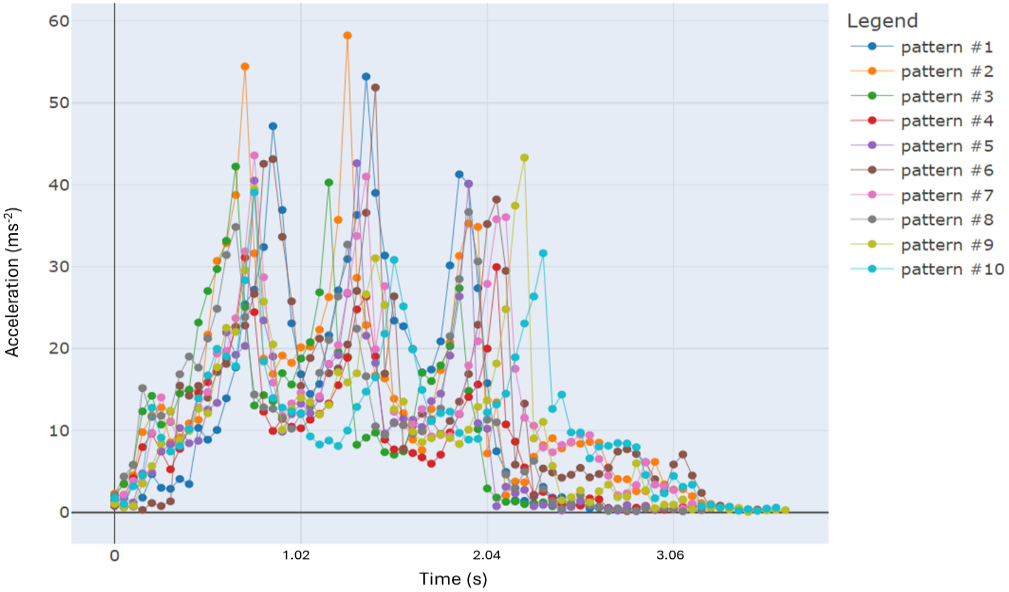}
         \caption{}
         \label{fig_participant2_accR}
     \end{subfigure}
     \caption{Acceleration readings of Participant 2 in: (a) x-axis (b) y-axis (c) z-axis (d) resultant vector (an aggregated data of the readings along the three axes)}
     \label{fig_participant2_acc}
\end{figure*}

\begin{figure*}
     \centering
     \begin{subfigure}{0.49\textwidth}
         \centering
         \includegraphics[width=\linewidth]{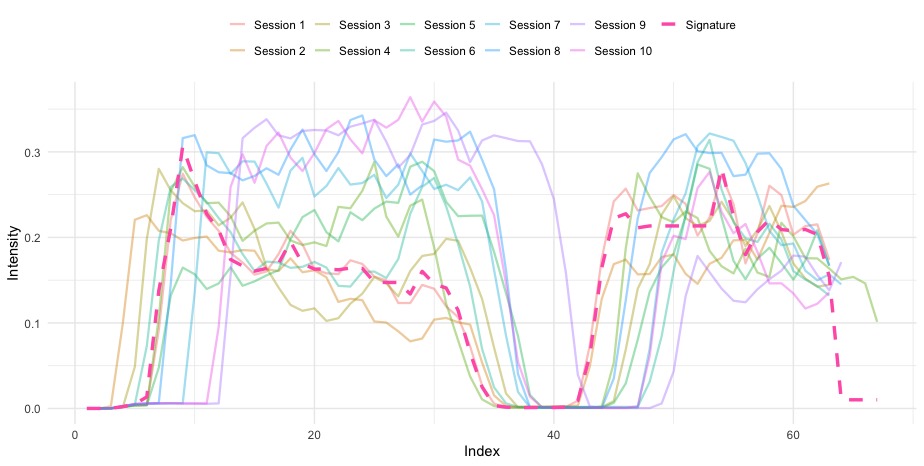}
         \caption{}
         \label{fig_participant1_r}
     \end{subfigure}
     \begin{subfigure}{0.49\textwidth}
         \centering
         \includegraphics[width=\linewidth]{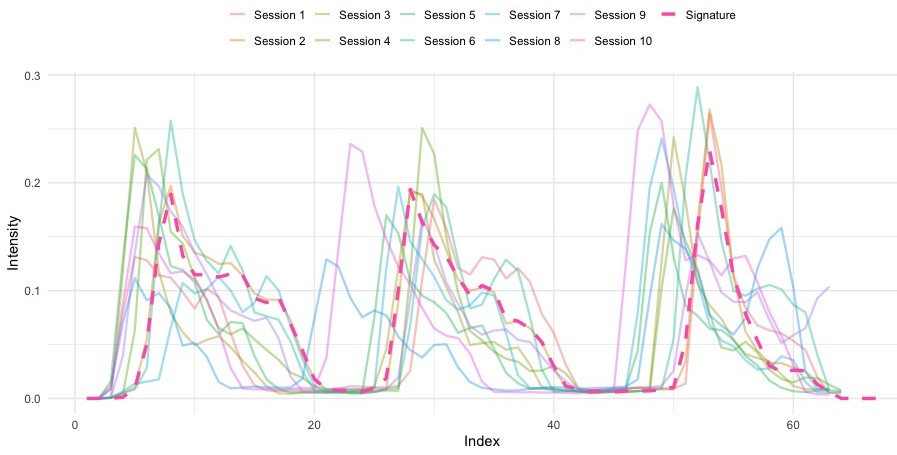}
         \caption{}
         \label{fig_participant2_r}
     \end{subfigure}
      \begin{subfigure}{0.49\textwidth}
         \centering
         \includegraphics[width=\linewidth]{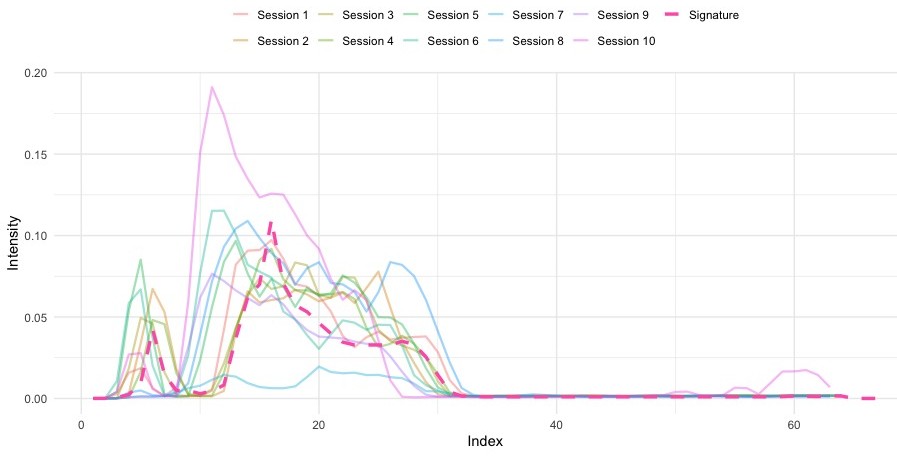}
         \caption{}
         \label{fig_participant3_3}
     \end{subfigure}
     \begin{subfigure}{0.49\textwidth}
         \centering
         \includegraphics[width=\linewidth]{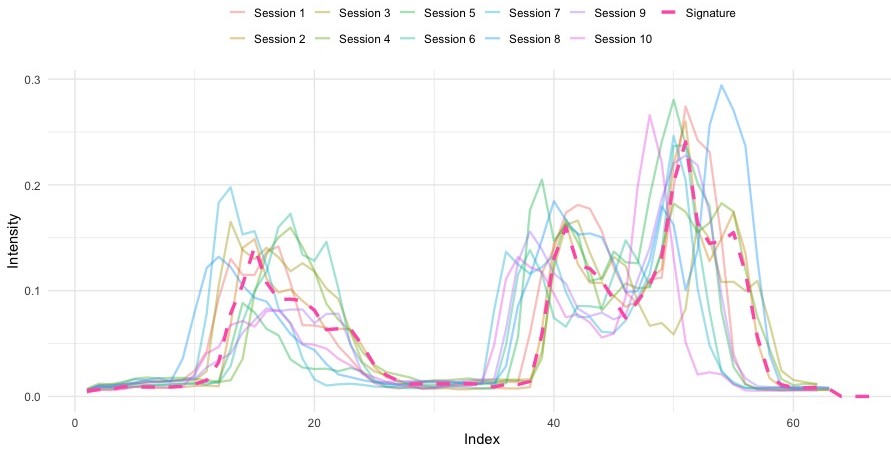}
         \caption{}
         \label{fig_participant4}
     \end{subfigure}
     \caption{Sample blow acoustic time series data of (a) Participant 1 (b) Participant 2 (c) Participant 3 (d) Participant 4, where ``Signature'' is an aggregate of the 10 sessions}
     \label{fig_sample_blow_acoustic_data}
\end{figure*}

\end{document}